\documentclass[aps,pra,twocolumn,amsmath,amssymb]{revtex4}
\usepackage{bm}
\usepackage{graphicx}

\newcommand{\be}{\begin{eqnarray}}
\newcommand{\ee}{\end{eqnarray}}

\newcommand{\ba}{\begin{array}}
\newcommand{\ea}{\end{array}}

\begin{document}

\title{Entanglement Assisted Random Access Codes}

\author{Marcin Paw{\l}owski}
\affiliation{Institute of Theoretical Physics and Astrophysics,
University of Gda\'nsk, 80-952 Gda\'nsk, Poland}

\author{Marek \.Zukowski}
\affiliation{Institute of Theoretical Physics and Astrophysics,
University of Gda\'nsk, 80-952 Gda\'nsk, Poland}

\begin{abstract}
An $(n,m,p)$ Random Access Code (RAC) allows to encode $n$ bits in an $m$ bit message, in such a way that a receiver of the message  can guess any of the original $n$ bits with probability $p$, greater than $\frac{1}{2}$. In Quantum RAC's (QRACs)  one transmits  $n$ qubits. The full set of primitive Entanglement Assisted Random Access Codes (EARACs) is introduced, in which parties are allowed to share a two-qubit singlet. It is shown that via a concatenation of these, one can build for any $n$ an $(n,1,p)$ EARAC.  QRAC's for $n>3$ exist only if parties additionally share classical randomness (SR). We show that  EARACs outperform the best of known QRACs not only in the success probabilities but also in the amount of communication needed in the preparatory stage of the protocol. Upper bounds on the performance of EARACs are given, and shown to limit also QRACs.
\end{abstract}

\maketitle

\section{Introduction}

In many communication related tasks quantum protocols are superior to classical ones. In cryptography, secret sharing and communication complexity two particular approaches to "quantization" of classical tasks are used. The parties use quantum communication instead of classical one, or the communication stays classical but the parties are allowed to share some entanglement. Examples for both approaches include \cite{BB84} and \cite{E91} in cryptography, \cite{qSS} and \cite{eSS} in secret sharing, \cite{qCC} and \cite{eCC} in communication complexity.

Quantum Random Access Codes (QRACs), since their introduction in \cite{Am1}, have been  studied only as protocols with quantum communication. Recently an experimental realization of QRAC's has been demonstrated \cite{QRACexp}. As RACs are extremely useful in information processing tasks (f.e. network coding \cite{RACapp}), it is important to search for optimal codes.   We address the problem, whether protocols that involve classical communication and entanglement lead to better codes. The answer is  positive.

\section{Primitives}
 
Every RAC is described by three numbers $n$, $m$ and $p$, where $n$ is the number of bits $(a_0,a_1,..,a_{n-1})$ that are known only by the first party (Alice),  and $m$ is the number of bits she sends to the second party (Bob), $m<n$. The code is optimized in such a way that for every bit $a_i$ of Alice, the probability that Bob correctly guesses this bit is at least $p$. Such code is denoted by $(n,m,p)$ or, in the case when the probability is not specified, but is strictly greater than $\frac{1}{2}$, by $(n,m)$. In the case of a QRAC the communicated bits are replaced by qubits. Entanglement Assisted Random Access Code (EARAC) is a code in which the $m$ communicated bits are classical, however the parties are additionally allowed to use shared entangled states during their coding/decoding procedures.
For a specified $m$ the only figure of merit to be studied is $p$ as a function of $n$ \footnote{We study the case of $m=1$. A generalization to $m>1$ will be presented elsewhere \cite{mg1}.}. Other parameters, like the amount of entanglement necessary for the code, are usually not taken into account. However, if we were to introduce some additional efficiency factor that quantifies the amount of communication required in the preparatory stage of the protocol, EARACs have the upper hand also in this area. To be able to compare different types of resources, we first need to equate a single bit of shared randomness (SR) with a single e-bit. This is justified, since both of these elementary resources can be generated by a transfer of a single qubit from one party to another. Now we note that for $(n,1)$ code, the EARACs presented in this paper require at most $n-1$ e-bits, while (for $n>3$) QRACs require at least $n$ bits of SR \cite{Am2}.

Let us start with pinpointing the two primitive EARACs, which are $(2,1,\frac{1}{2}\big( 1+\frac{1}{\sqrt{2}}\big))$ and $(3,1,\frac{1}{2}\big( 1+\frac{1}{\sqrt{3}}\big))$ ones, achievable with a shared two-qubit singlet, and without any shared classical randomness. In the first case, denoted as $E^{[2]}$, Alice encodes her bits two, $a_0$ and $a_1$, by making a measurement in a basis dependent on the value of $a_0 \oplus a_1$. The two bases of Alice $A_{a_0\oplus a_1}$, are specified by the following Bloch vector pairs
$
A_a=\{\pm\frac{1}{\sqrt{2}}(1,(-1)^a,0)\},
$
where $a=0,1$.
Alice's outcome $A$ is denoted as 0, if the measurement results in a collapse onto the first ``$+$'' state of the basis, and 1 if onto the second one. Whenever Bob wants to learn a $b$-th bit of Alice, he chooses to measure in the basis $B_b$, defined as
\be \label{basesBob1}
B_0=\{\pm(1,0,0)\}, \quad B_1=\{\pm(0,1,0)\} .
\ee
Bob's outcome $B$ is ascribed 0 for the $-$ vectors and $1$ for $+$ ones. Alice sends to Bob a message $M=a_0\oplus A$. With this he can decode the desired bit with a high probability. As $P(A\oplus B=0)=\frac{1}{2}(1+\vec{a}\cdot\vec{b})$, where $\vec{a}$ and $\vec{b}$ are the Bloch vectors of the local settings specified by the $+$ vectors of the bases, which in turn implies that $P(A\oplus B=ab)=\frac{1}{2}\big( 1+\frac{1}{\sqrt{2}}\big)$, it is easy to check that
\be
M\oplus B= a_b
\ee
with a probability $p=\frac{1}{2}\big( 1+\frac{1}{\sqrt{2}}\big)\approx 0.85$. This code may be considered as version of an Oblivious Transfer protocol presented in \cite{WW} with the singlet state being an imperfect realization of a PR Box \cite{PR}.

To construct a $(3,1,\frac{1}{2}\big( 1+\frac{1}{\sqrt{3}}\big))$ EARAC, denoted here as $E^{[3]}$, a similar procedure can be used. The only difference is in the bases that the parties choose. For Alice there are 4 bases and the choice is again implied by her input bits
\be \nonumber
a_0=a_1=a_2 &\Rightarrow& A_0=\{\pm\frac{1}{\sqrt{3}}(1,1,1)\},
\\ \nonumber
a_0=a_1\neq a_2 &\Rightarrow& A_1=\{\pm\frac{1}{\sqrt{3}}(1,1,-1)\},
\\ \nonumber
a_0\neq a_1=a_2 &\Rightarrow& A_2=\{\pm\frac{1}{\sqrt{3}}(1,-1,-1)\},
\\
a_0\neq a_1 \neq a_2 &\Rightarrow& A_3=\{\pm\frac{1}{\sqrt{3}}(1,-1,1)\}.
\ee
The outcome and the message are defined exactly in the same way as in the previous $(2,1)$ code. Bob again uses mutually unbiased bases $B_0$ and $B_1$ from (\ref{basesBob1}) but this time, if he aims at the third bit, he  chooses
$
B_2=\{\pm(0,0,1)\}.
$
It is easy to check, that in this case the probability that $M\oplus B=a_b$ is $p=\frac{1}{2}\big(1+\frac{1}{\sqrt{3}}\big)\approx 0.79$. To our knowledge this protocol has not been studied so far.

The success probabilities for these codes exactly match the ones for known QRACs \cite{Am1,RACapp}. This is not surprising. The basis vectors of Alice in EARAC correspond exactly to her quantum codewords in QRAC, and Bob's measurements are the same.

In \cite{RACapp} it was shown that a $(4,1)$ QRAC (without shared randomness) does not exist (this also holds for any $(n,1)$ QRAC with $n > 4$). Nevertheless, as we will show for any $n$ one has an $(n,1)$ EARAC. To this end we present concatenated EARACs formed out of the primitive ones singled out above.

\section{Concatenation}

Since the  procedure has classical inputs and outputs at every point it can be concatenated.  Consider the following  simplest case. Alice is given 4 bits $a_0,a_1,a_2$ and $a_3$, while Bob might be interested in a bit number $b=0,...,3$.
They agree that Bob would use a binary expansion of $b$, given by a bit sequence $b_1b_0$ defined via  $\sum_{i=0}^1 b_i2^i$, where $b_i=0,1$.
Alice encodes the first two bits of hers by performing a measurement in a basis $A_{a_0\oplus a_1}$  on a $(2,1)$ EARAC called here $E(b_1=0)$. Denote the measurement result  by $A^{(0)}$. Her output value is $M_0=a_0\oplus A^{(0)}$.  She also encodes the other two bits in a similar procedure with yet another $(2,1)$ EARAC, denoted as $E(b_1=1)$. The basis is now  $A_{a_2\oplus a_3}$. The output is $M_1=a_2\oplus A^{(1)}$, where $A^{(1)}$ is her measurement result for $E(b_1=1)$. The value of $M_1$ along with $M_0$ are treated as her input bits for a third $(2,1)$ EARAC, denoted by $E'$, which fixes the measurement basis as $A_{M_0\oplus M_1}$. Let us denote the result by $A'$. Only the output bit of this final EARAC, $M=M_0\oplus A'$ is then sent to Bob.
What Alice does is an encoding of  $a_0$ and $a_1$ into $M_0$ and  $a_2$ and $a_3$ into $M_1$, while $M_0$ and $M_1$ are next
encoded into $M$. When Bob gets $M$, and performs a measurement on $E'$ in the basis $B_{b_1}$ yielding a result denoted as $B'$, he is able to decode $M_{b_1}$,  if he  also performs a  measurement in basis $B_{b_0}$ on $E(b_1)$. The results for $E(b_1=x)$ we denote by $B^{(x)}$.  The value of his guess is then: $a_{b_1b_0}=M\oplus B^{(b_1)}\oplus B'$. Bob makes measurements on only two out of three singlets in his disposal, but which of them are measured depends on his choice of bit he is interested in. Once he chooses to measure $E'$ in the basis $B_{b_1=x}$, this can be accompanied by a measurement on $E(b_1=x)$, a measurement on $E(b_l\neq x)$ is useless.  A larger number of bits encoded comes at the price of lower probability of success. Bob will guess the target bit correctly if both of their EARAC devices give the correct value, or if both of them are wrong. Therefore, the code just introduced is a $(4,1,\frac{3}{4})$ EARAC (see below).

More generally, if Alice and Bob know how to devise a $(k,1,p_k)$ EARAC, denoted here $E^{[k]}$, they can use two such procedures followed by with a $(2,1)$ one to construct a $(2k,1,p_{2k})$ code. Alice  simply encodes first $k$ bits into one the first $E^{[k]}$ using the coding procedure from $(k,1,p_k)$ code and she repeats the procedure for the remaining $k$ bits with the second $E^{[k]}$. That allows her to compress $2k$ bits into two,  which she can encode using a $(2,1)$ primitive EARAC into a single bit message, which is send to Bob. Again, Bob needs two failures or two successes of their EARAC devices to get the correct value of the bit he is interested in. This yields the probability of success
\be
p_{2k}=p_k\frac{1}{2}\big( 1+\frac{1}{\sqrt{2}}\big)+(1-p_k)\frac{1}{2}\big( 1-\frac{1}{\sqrt{2}}\big),
\ee
which if one puts $p_k=\frac{1}{2}(1+d)$ reads $p_{2k}=\frac{1}{2}(1+d\frac{1}{\sqrt{2}})$
By induction, since $p_1=\frac{1}{2}\big( 1+\frac{1}{\sqrt{2}}\big)$, this procedure allows to generate a $(2^k,1,\frac{1}{2}(1+2^{-\frac{1}{2}k}))$ EARAC.

To generate an $(n,1)$ EARAC for  $n>3$ one must use the primitive ones, $E^{[2]}$ and $E^{[3]}$, in a concatenated scheme, see Fig.1. For example for $n=5$ Alice will encode first three bits using a $E^{[3]}$ and the remaining two with $E^{[2]}$. Then she uses one more $E^{[2]}$ to get one bit of message that she sends to Bob. In  the general case, if Alice gets $n$ bits she can divide them in $n_2$ groups of 2 and $n_3$ groups of 3. If the reminder of the division of $n$ by 3 is $r$ then $n_2=2r \mod 3$ and $n_3=n-2n_2$. This gives her $n'=\frac{n_3}{3}+\frac{n_2}{2}\geq n$ bits, with which she will perform a similar procedure until she is left with only 1 bit, that she can communicate to Bob (of course they ignore the possible additional bit, by fixing it to $0$).

To calculate the success probability of guessing a given bit one has to trace the way from the initial bit to the final coded message in the schematic representation of the code (see Fig. 1 for example). Since the success probability for a $E^{[3]}$ is lower then for $E^{[2]}$, the success probability of the code will depend on which primitives were used in the concatenation. The guess is be successful not only when the guess is correct for every step of the concatenation, but also if the number of errors is even. The probability to make an even number of errors when using  $E^{[2]}$ $k$ times is
\begin{eqnarray} &
p_{2e}(k)=\sum_{i=0}^{\lfloor\frac{k}{2}\rfloor}{k\choose 2i}\Big(\frac{1}{2}\big( 1+\frac{1}{\sqrt{2}}\big) \Big)^{k-2i}
\Big(\frac{1}{2}\big( 1-\frac{1}{\sqrt{2}}\big) \Big)^{2i}&\nonumber
\\
&=\frac{1}{2}\big(1+2^{-\frac{1}{2}k}\big).&
\end{eqnarray}
For a $E^{[3]}$ it is
\be
p_{3e}(k)=\frac{1}{2}\big(1+3^{-\frac{1}{2}k}\big).
\ee
For an odd number of errors one has
\be
p_{2o}(k)=\frac{1}{2}\big(1-2^{-\frac{1}{2}k}\big),
\\
p_{3o}(k)=\frac{1}{2}\big(1-3^{-\frac{1}{2}k}\big).
\ee
That gives the overall probability of success for a bit encoded $k$ times with $(2,1)$ EARACs and $j$ times with $(3,1)$ EARACs:
\be \nonumber
p_{k,j}=p_{2e}(k)p_{3e}(j)+p_{2o}(k)p_{3o}(j)=\frac{1}{2}\big(1+2^{-\frac{1}{2}k}3^{-\frac{1}{2}j}\big). \\ \label{og}
\ee
Thus, for any encoding of this type the probability to guess any bit is strictly greater than $\frac{1}{2}$. Therefore,  in contrast to QRACs, an $(n,1)$ EARAC exists for any $n$, even if the parties do not make use of shared randomness.

When parties are allowed to use shared randomness (SR),  QRAC do exist for any $n$, see \cite{Am2}.  Obviously, for any one can devise non-concatenated EARAC using in the same way SR,  as their counterpart QRACs.  However, we will show that for concatenated EARAC with SR, the success probabilities are higher than for the best of known QRAC's.

\begin{figure}[ht]
\includegraphics[scale=1]{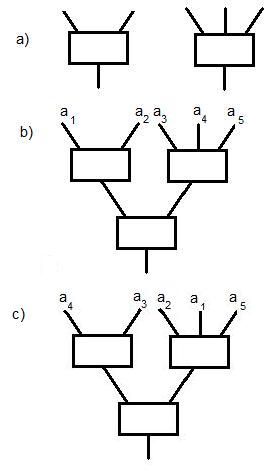}
  \caption{a) Schematic representation of the two primitives: $(2,1)$ and $(3,1)$ EARAC; b) Example of concatenation: $(5,1)$ EARAC. The probability for Bob to guess $a_2$ is higher than for $a_3$ since in the first case only $(2,1)$ EARAC primitives are used. These probabilities are respectively $P_2=\frac{1}{2}\big( 1+\frac{1}{\sqrt{4}}\big)$ and $P_3=\frac{1}{2}\big( 1+\frac{1}{\sqrt{6}}\big)$. The success probability associated with the whole code is the lower one; c) One of possible permutation of input bits for $(5,1)$ code. The choice of permutation is governed by the data from shared random string. Having all possible permutations with the same probability results in an averaged version of the protocol with the success probability $p=\frac{2}{5}P_2+\frac{3}{5}P_3$, which is higher than in the previous case.}
\label{codes}
\end{figure}

\section{Shared randomness}

One of the possibilities of employing SR in an EARAC is via defining the place each bit of Alice $a_i$ enters the coding procedure by the values of the random string. This allows to average the success probabilities. Obviously this cannot decrease the minimal probability, which defines the efficiency of the code. In fact in most of the cases this efficiency is increased. For an $(n,1,p)$ EARAC with $n_i$ bits having success probability $p_i$, where $n=\sum_i n_i$ and $p=\min p_i$, SR allows an upgrade to an $(n,1,p')$ EARAC, where
$
p'=\frac{1}{n}{\sum_i n_i p_i}
$
(see Fig. 1).

An evident advantage of an EARAC with SR over a QRAC with SR from \cite{Am2}, which are the best ones of the known so far, is a simpler way to find a construction. To create the EARAC one just needs to concatenate the two basic codes a sufficient number of times. In the case of a QRAC numerical search procedures are used. The other advantage is that EARACs give higher probabilities of success for all $n>3$, see   Table 1.
\begin{table}
\begin{tabular}{|c|c|c|c|}
\hline
$n$ &  $p_{Q,n}$ & $p_{E,n}$ & $\Delta=p_{E,n}-p_{Q,n}$ \\ \hline
2 & $\frac{1}{2}\big(1+\frac{1}{\sqrt{2}}\big)$ & $\frac{1}{2}\big(1+\frac{1}{\sqrt{2}}\big)$ & 0  \\ \hline
3 & $\frac{1}{2}\big(1+\frac{1}{\sqrt{3}}\big)$ & $\frac{1}{2}\big(1+\frac{1}{\sqrt{3}}\big)$ & 0  \\ \hline
4 & 0.74148 & $\frac{3}{4}$ & 0.00852 \\ \hline
5 & 0.71358 & $\frac{1}{20}(12+\sqrt{6})$ & 0.00889 \\ \hline
6 & 0.69405 & $\frac{1}{2}\big(1+\frac{1}{\sqrt{6}}\big)$ & 0.01007 \\ \hline
7 & 0.67864 & $\frac{1}{21}(12+\sqrt{6})$ & 0.00943 \\ \hline
8 & 0.66663 & $\frac{1}{80}(52+\sqrt{6})$ & 0.01399 \\ \hline
9 & 0.65689 & $\frac{2}{3}$ & 0.00978 \\ \hline
10 & 0.64820 & $\frac{1}{20}(10+\sqrt{2}+\sqrt{3})$ & 0.00911 \\ \hline
11 & 0.64105 & $\frac{1}{120}(60+3\sqrt{2}+8\sqrt{3})$ & 0.00978 \\ \hline
12 & 0.63487 & $\frac{1}{2}\big(1+\frac{1}{\sqrt{12}}\big)$ & 0.00947 \\ \hline
15 & 0.62036 & $\frac{1}{60}(30+3\sqrt{2}+2\sqrt{3})$  & 0.00809 \\ \hline
\end{tabular}
\caption{Success probabilities for $(n,1,p_{Q,n})$ QRAC with SR from \cite{Am2} compared with the probabilities for
$(n,1,p_{E,n})$ EARAC. Last column displays the advantage of EARAC for every $n>3$.}
\end{table}

\section{Upper and lower bound}

Calculating the exact success probability for $(n,1,p)$ EARAC is not difficult, but hard to be put in a  short formula.
Nevertheless, it is possible in the special case when the success probabilities for all individual bits are equal, without a randomization procedure with the use of SR. If  $k$ and $j$ again denote the number of times $(2,1)$ and $(3,1)$ are used for encoding a bit, $p_{k,j}$ will be the same for all bits if and only if for {\em all} bits $k$ and $j$ are the same. This implies that the above conditions hold for only
$
n=2^k 3^j.
$
In such case,  via Eq. (\ref{og}), the success probability is
\be \label{pn}
p=p_n=\frac{1}{2}\big(1+\frac{1}{\sqrt{n}}\big).
\ee

Equation (\ref{pn}) gives also an upper bound for the success probability of any EARAC for any $n$. Consider an $(n^N,1)$ EARAC constructed by $N$-level concatenation of  $(n,1,\frac{1}{2}\big(1+\frac{1}{\sqrt{n}}+\epsilon\big))$ EARACs. Since Bob will guess every bit in the concatenated protocol with the same probability, the success probability is given by (\ref{pn})
\be
p_{n^N}=\frac{1}{2}\Bigg(1+\Big(\frac{1}{\sqrt{n}}+\epsilon\Big)^N\Bigg).
\ee
To proceed further we shall use the following technical result.
In \cite{IC} it has been shown that, in the general case on any information processing protocol, classical or quantum, if Alice holds  a $K$-bit secret entirely random data string and sends {\em one} classical bit to Bob, then the bound on mutual information implies the following inequality
\be \label{ic}
K(1-h(p_K))\leq 1
\ee
where $p_K$ denotes the probability that Bob guesses correctly any of her $K$ bits, and $h$ is Shannon binary entropy. If we substitute $K=n^N$ and use $1-h\left(\frac{1+y}{2}\right)\ge\frac{y^2}{2\ln 2}$ we get
\be \nonumber
K(1-h(p_K))\geq \frac{n^N\Big(\frac{1}{\sqrt{n}}+\epsilon\Big)^{2N}}{2\ln 2}= \frac{\Big(1+2\sqrt{n}\epsilon+n\epsilon^2\Big)^{N}}{2\ln 2}.\\
\ee
If $\epsilon>0$ then $1+2\sqrt{n}\epsilon+n\epsilon^2>1$ and there exists $N$ large enough for which
$\Big(1+2\sqrt{n}\epsilon+n\epsilon^2\Big)^{N}>2\ln 2$ which leads to violation of (\ref{ic}). That means that for any $(n,1,p_n)$ EARAC one must have
\be \label{bound}
p_n\leq \frac{1}{2}\big(1+\frac{1}{\sqrt{n}}\big).
\ee
Note that the bound (\ref{ic}) also holds for QRACs. To see this, notice that for any $(n,1,p)$ QRAC there exists $(n,1,p)$ EARAC with the same $p$. It can be constructed in the following way. Alice and Bob share a singlet state. When Alice wants to encode her input $a_1,..,a_n$ she calculates $\rho(a_1,..,a_n)$, which would be the (pure) state that she would send to Bob if they were to use $(n,1,p)$ QRAC. She then measures her part of the singlet in the basis that includes $\rho(a_1,..,a_n)$ as one of its eigenstates. Her measurement outcome tells her whether the part of the singlet at Bob's lab collapsed to $\rho(a_1,..,a_n)$ or the orthogonal one. She then sends Bob 0 if it is this state and 1 if it is orthogonal. When Bob wants to find the value of $i$-th bit he performs the measurement that he would in the case of QRAC. If he has  received message 0 from Alice he keeps his outcome, if 1 he flips it. That procedure gives him the same probability of successfully guessing any bit as in QRAC. Therefore any bound on such EARACs is also valid for QRACs. Our proof is thus also a simpler version of the one for the bound for QRAC presented in \cite{Am2}. The bound (\ref{bound}) is saturated whenever $n$ is of the type $2^k3^j$. Thus,  in such cases the presented EARACs are optimal.

As it has already been mentioned, the derivation of success probability for any given EARAC is straightforward, but it is difficult to give one formula for the general case. It is however possible to give a lower bound for optimal protocols. Notice that if $n$ is not of the form $2^k3^j$ the parties can always use an $(n_\geq,1)$ EARAC, where $n_\geq$ is the smallest integer greater than $n$ and being of the form $2^k3^j$. This leads to the lower bound
\be
p_n\geq \frac{1}{2}\big(1+\frac{1}{\sqrt{n_\geq}}\big).
\ee

\section{Conclusion}

We introduce Entanglement Assisted Random Access Codes and  show them to be superior to the best known QRACs in the terms of both success probabilities, and existence without SR. We  also derive the bound for any Random Access Code and show that for infinitely many $n$ the EARAC is the best one. An open questions is whether the bound is saturated for all $n$, and if not, what are then the best possible codes.

\acknowledgements

We thank T. Paterek for discussions. This work has been supported by EU programme QAP (no. 015848). It has been done at the National Quantum Information Centre of Gdansk (NQuantIC).

\end{document}